\documentclass[aps,twocolumn,showpacs,preprintnumbers,amsmath,amssymb]{revtex4}

\usepackage{graphicx}
\usepackage{dcolumn}
\usepackage{bm}

\begin{document}

\preprint{Preprint: Final}

\title{Optimal Sizes of Dielectric Microspheres for Cavity QED with Strong Coupling}
\author{J. R. Buck and H. J. Kimble}
\affiliation{Norman Bridge Laboratory of Physics 12-33,\\
California Institute of Technology, Pasadena, CA 91125}
\date{October 5, 2002}

\begin{abstract}
The whispering gallery modes (WGMs) of quartz microspheres are
investigated for the purpose of strong coupling between single
photons and atoms in cavity quantum electrodynamics (cavity QED).
Within our current understanding of the loss mechanisms of the
WGMs, the saturation photon number, $n_{0}$, and critical atom
number, $N_{0}$, cannot be minimized simultaneously, so that an
\textquotedblleft optimal\textquotedblright\ sphere size is taken
to be the radius for which the geometric mean, $\sqrt{n_{0}\times
N_{0}}$, is minimized. While a general treatment is given for the
dimensionless parameters used to characterize the atom-cavity
system, detailed consideration is given to the $D_2$ transition in
atomic Cesium at $\lambda_{0} =852\mathrm{nm}$ using fused-silica
microspheres, for which the maximum coupling coefficient
$g_{a}/(2\pi )\approx 750\mathrm{MHz}$ occurs for a sphere radius
$a=3.63\mathrm{\mu m}$ corresponding to the minimum for
$n_{0}\approx 6.06\times 10^{-6}$. By contrast, the minimum for
$N_{0}\approx 9.00\times 10^{-6}$ occurs for a sphere radius of
$a=8.12\mathrm{\mu m}$, while the optimal sphere size for which
$\sqrt{n_{0}\times N_{0}}$ is minimized occurs at
$a=7.83\mathrm{\mu m}$. On an experimental front, we have
fabricated fused-silica microspheres with radii $a \sim
10\mathrm{\mu m}$ and consistently observed quality factors $Q\geq
0.8\times 10^{7}$. These results for the WGMs are compared with
corresponding parameters achieved in Fabry-Perot cavities to
demonstrate the significant potential of microspheres as a tool
for cavity QED with strong coupling.
\end{abstract}


\maketitle

\section{Introduction}

Motivated by the pioneering work of Braginsky and Ilchenko
\cite{braginsky1987}, some of the highest quality optical
resonators to date have been achieved with the whispering gallery
modes (WGMs) of quartz microspheres \cite{gorodetsky,vernooy2}.
Over the wavelength range $630-850\mathrm{nm}$, quality factors
$\protect{Q\approx 8\times 10^{9}}$ have been realized, and cavity
finesse $\mathcal{F}=2.3\times 10^{6}$ demonstrated
\cite{gorodetsky,vernooy2}. Such high quality factors make the
WGMs of small dielectric spheres a natural candidate for use in
cavity QED \cite{vernooy1, mabuchi, braginsky1987, gorodetsky2,
haroche1994, haroche1995, haroche1996, jhe, chew, klimov, klimov2,
ilchenko2001, chang, campillo, slusher,matsko,moerner,wang}.

While much of the work regarding quartz microspheres has centered
around achieving the ultimate quality factors
\cite{gorodetsky,vernooy2}, the quality factor of the resonator is
but one of the factors that determines the suitability of the WGMs
for investigations of cavity quantum electrodynamics in a regime
of strong coupling. In this case, the coherent coupling
coefficient, $g$, for a single atom interacting with the cavity
mode must be much larger than all other dissipative rates,
including the cavity decay rate, $\kappa$, and the rate of atomic
spontaneous emission, $\gamma$; namely $g\gg (\kappa ,\gamma )$.
Note that $2g=\Omega$ gives the Rabi frequency associated with a
single quantum of excitation shared by the atom-cavity system
\cite{jeff,jeff2}. The atom-field interaction can be characterized
by two important dimensionless parameters: the saturation photon
number, $n_{0}\propto \frac{\gamma^{2}}{g^{2}}$, and the critical
atom number, $N_{0}\propto\frac{\kappa\gamma}{g^{2}}$. Since these
parameters correspond respectively to the number of photons
required to saturate an intracavity atom and the number of atoms
required to have an appreciable effect on the intracavity field,
strong coupling requires that $(n_{0},N_{0})\ll 1$. Ideally one
would hope to minimize both of these parameters in any particular
resonator. Unfortunately, within the context of our current
understanding of the loss mechanisms of the WGMs \cite{vernooy2},
the critical parameters $(n_{0},N_{0})$ cannot be minimized
simultaneously in a microsphere.

Motivated by these considerations, in this paper we explore
possible limits for the critical parameters $(n_{0},N_{0})$ for
the WGMs of quartz microspheres. Following the analysis of Refs.
\cite{mabuchi, vernooy1, matsko}, we study the particular case of
a single atom coupled to the \textit{external} field of a WGM near
the sphere's surface. We show that there are radii that minimize
$(n_{0},N_{0})$ individually, and that there is an
\textquotedblleft optimal" sphere size that minimizes the
geometric mean, $\sqrt{n_{0}\times N_{0}}$, of these two
cavity-QED parameters and allows both parameters to be near their
respective minima. We also report our progress in the fabrication
of small microspheres with radii $a\sim 10\mathrm{\mu m}$, and
compare our experimental results for $Q$ with those from our
theoretical analysis. Finally, we present a detailed comparison
for the state of the art and future prospects for achieving strong
coupling in cavity QED for both microsphere and Fabry-Perot
cavities. Throughout the presentation, we attempt to develop a
general formalism that can be applied to diverse systems. However,
for definiteness we also present results for a particular system
of some interest, namely an individual Cesium atom coupled to the
WGMs of quartz microspheres.

\section{Modes of a Microsphere}

Solving for the mode structure of the resonances of a dielectric
sphere in vacuum is a classic problem in electricity and
magnetism, and the resulting field distributions have been known
for some time \cite{stratton}. The electric field of the TM,
\textsl{electric type\/}, modes inside and outside a sphere of
refractive index $n$ at free-space wavelength $\lambda_{0}$ are
respectively,
\begin{eqnarray}
\vec{E}_{{\rm in}}(r,\theta,\phi) \propto && l(l + 1)\frac{j_{l}(k
r)}{k
r}P_{l}^{m}(\cos\theta)e^{i m \phi}\widehat{r}\nonumber\\
    + && \frac{\left[k r j_{l}(k r)\right]'}{k r}\frac{\partial P_{l}^{m}(\cos\theta)}{\partial\theta}
    e^{i m \phi}\widehat{\theta}\nonumber\\
    + && \frac{i m}{\sin \theta}\frac{\left[k r j_{l}(k r)\right]'}{k r}P_{l}^{m}(\cos \theta)
    e^{i m \phi}\widehat{\phi}
\end{eqnarray}
and,
\begin{eqnarray}
\vec{E}_{{\rm out}}(r,\theta,\phi) \propto && l(l +
1)\frac{h^{(1)}_{l}\left(\frac{k r}{n}\right)}
    {\frac{k r}{n}}P_{l}^{m}(\cos\theta)e^{i m\phi}\widehat{r}\nonumber\\
    + && \frac{\left[\frac{k r}{n} h^{(1)}_{l}\left(\frac{k r}{n}\right)\right]'}{\frac{k r}{n}}
    \frac{\partial P_{l}^{m}(\cos\theta)}{\partial\theta}e^{i m \phi}\widehat{\theta}\nonumber\\
    + && \frac{i m}{\sin \theta}\frac{\left[\frac{k r}{n} h^{(1)}_{l}\left(\frac{k r}{n}\right)\right]'}{\frac{k r}{n}}
    P_{l}^{m}(\cos \theta)e^{i m \phi}\widehat{\phi}\;.
\end{eqnarray}
where $a$ is the radius of the sphere, $k=\frac{2\pi
n}{\lambda_{0}}$ is the wave vector inside the sphere, $j_{l}(x)$
is the spherical Bessel function, $h_{l}^{(1)}(x)$ is the
spherical Hankel function,
$(\widehat{r},\widehat{\theta},\widehat{\phi})$ are unit vectors,
and the $'$ refers to differentiation with respect to the
argument. Note that the TM modes have a predominantly radial
electric field vector.

\begin{figure}[tb]
\includegraphics[width=8.6cm]{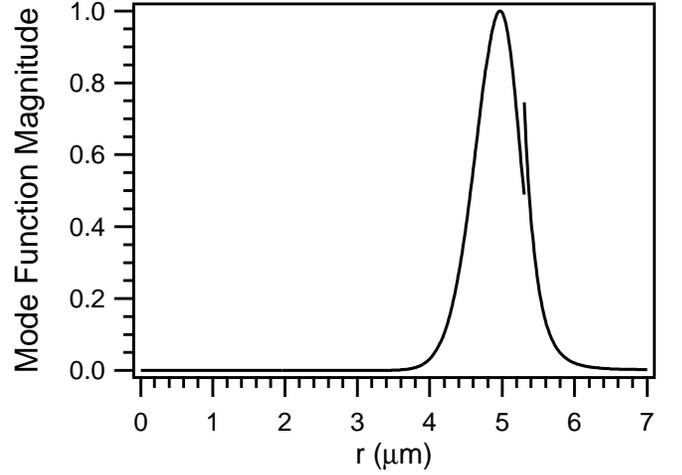}
\caption{\label{mode-function}The magnitude of the normalized mode
function as a function of radius for the TM mode of a
$5.305\mathrm{\protect\mu m}$ radius sphere ($p=1,l=m=50$) with
$\protect\theta =\frac{\protect\pi}{2}$ and $\protect\phi =0$ for
a wavelength of $\lambda_{0} =852.359\mathrm{nm}$ and index of
refraction $n=1.45246$. In our case, the function is normalized to
have a maximum value of unity. Note that there is a discontinuity
at the surface.}
\end{figure}

In order to satisfy the boundary conditions at the surface of the
microsphere, the tangential components of the mode function
immediately inside and outside the sphere must be equal. However,
there is a discontinuity in the radial component of the electric
field at the dielectric boundary (as can be seen from
Fig.~\ref{mode-function}.) The eigenmodes are determined by
solving for the roots of a characteristic equation
\cite{stratton}, which can be reduced to
\begin{equation}
\frac{j_{l-1}(k a)}{j_{l}(k a)} - \frac{n
h_{l-1}^{(1)}\left(\frac{k a}{n}\right)}{h_{l}^{(1)}\left(\frac{k
a}{n}\right)}
 + \frac{n^{2} l}{k a} - \frac{l}{k a} = 0\;.
\label{characteristic}
\end{equation}

Throughout this paper, we normalize the mode functions such that
their maximum value is unity. This condition then yields for the
$l=m$ modes of the sphere
\begin{eqnarray}
\vec{\Psi}_{{\rm in}}(r,\theta,\phi) = && N (l + 1)\frac{j_{l}(k
r)}{k r}\sin^{l}(\theta)e^{i l \phi}\widehat{r}\nonumber\\
    + && N F(r) \cos\theta \sin^{l-1}\theta e^{i l \phi}\widehat{\theta}\nonumber\\
    + && i N F(r) \sin^{l-1}\theta e^{i l \phi}\widehat{\phi}
\end{eqnarray}
and,
\begin{eqnarray}
\vec{\Psi}_{{\rm out}}(r,\theta,\phi) = && N B (l +
1)\frac{h_{l}^{(1)}\left(\frac{k r}{n}\right)}{\frac{k
r}{n}}\sin^{l}\theta e^{i l \phi}\widehat{r}\nonumber\\
    + && N B H(r) \cos\theta \sin^{l-1}\theta e^{i l \phi}\widehat{\theta}\nonumber\\
    + && i N B H(r) \sin^{l-1}\theta e^{i l \phi}\widehat{\phi},
\end{eqnarray}
where
\begin{equation}
F(r) = \frac{j_{l}(k r)}{k r} + \frac{l}{2 l + 1}j_{l}(k r) -
\frac{l+1}{2l+1}j_{l+1}(k r),
\end{equation}
\begin{eqnarray}
H(r) = && \frac{h_{l}^{(1)}\left(\frac{k r}{n}\right)}{\frac{k
r}{n}}
    + \frac{l}{2l+1}h_{l-1}^{(1)}\left(\frac{k r}{n}\right)\nonumber\\
    &&- \frac{l+1}{2l+1}h_{l+1}^{(1)}\left(\frac{k r}{n}\right),
\end{eqnarray}
\begin{equation}
B = \frac{\frac{j_{l}(k a)}{k a} + \frac{l}{2 l + 1}j_{l}(k a) -
\frac{l+1}{2l+1}j_{l+1}(k a)}{\frac{h_{l}^{(1)}\left(\frac{k
a}{n}\right)}{\frac{k a}{n}}
    + \frac{l}{2l+1}h_{l-1}^{(1)}\left(\frac{k a}{n}\right)
    - \frac{l+1}{2l+1}h_{l+1}^{(1)}\left(\frac{k a}{n}\right)},
\end{equation}
and $N$ is the normalization factor. Because we will require the
field outside the sphere to be as large as possible, we will
choose the $p=1$ modes. Also, because the coherent coupling
constant $g\varpropto \frac{1}{\sqrt{V_{\vec{P}}}}$, where
$V_{\vec{P}}$ is the cavity mode volume, we choose the $l=m$
modes, since they yield the smallest electromagnetic mode volume,
as will be explained in the next section.

\section{Electromagnetic Mode Volume}

The effective mode volume $V_{\vec{P}}$ associated with the
electromagnetic field distribution $\vec{\Psi}(r,\theta ,\phi)$
\cite{vernooy1} is given by
\begin{equation}
V_{\vec{P}} =
\int_{V_{Q}}\varepsilon\left(\vec{r}\right)\left|\vec{\Psi}_{\vec{P}}
(\vec{r})\right|^{2} {\rm d}V\;,
\end{equation}
where
\begin{equation}
\varepsilon\left(\vec{r}\right) =
    \begin{cases}
        n^{2}& \text{if $r < a$},\\
        1& \text{if $r > a$}.
    \end{cases}
\end{equation}
and $\vec{P}$ corresponds to the $(p,l,m)$ mode. $V_{Q}$ is the
quantization volume discussed in Ref. \cite{vernooy1}. As long as
a radius $r_{Q}$ is chosen large enough to include the effects of
the evanescent field, the mode volume is relatively insensitive to
the particular choice of quantization radius \cite{r_qfootnote}.
As discussed more extensively in Refs. \cite{jeff, jeff2} the
interaction between the internal atomic degrees of freedom and the
intracavity field is characterized by the coherent coupling
constant $g(r,\theta ,\phi )$, where
\begin{equation}
g(r,\theta ,\phi )\equiv g_{0}\vec{\Psi}^{(p,l,m)}(r,\theta ,\phi
)
\end{equation}
and
\begin{equation}
g_{0}\varpropto \frac{1}{\sqrt{V_{\vec{P}}}}\text{.}
\label{geezero}
\end{equation}
Note that in the absence of damping, $2g\left(\vec{r}\right)$
gives the frequency for Rabi nutation associated with a single
photon in the cavity for an atom initially in the ground state
located at position $\vec{r}$ within the mode. Therefore, in order
to maximize the coupling strength, one must endeavor to minimize
the cavity mode volume.

\begin{figure}[tb!]
\includegraphics[width=8.6cm]{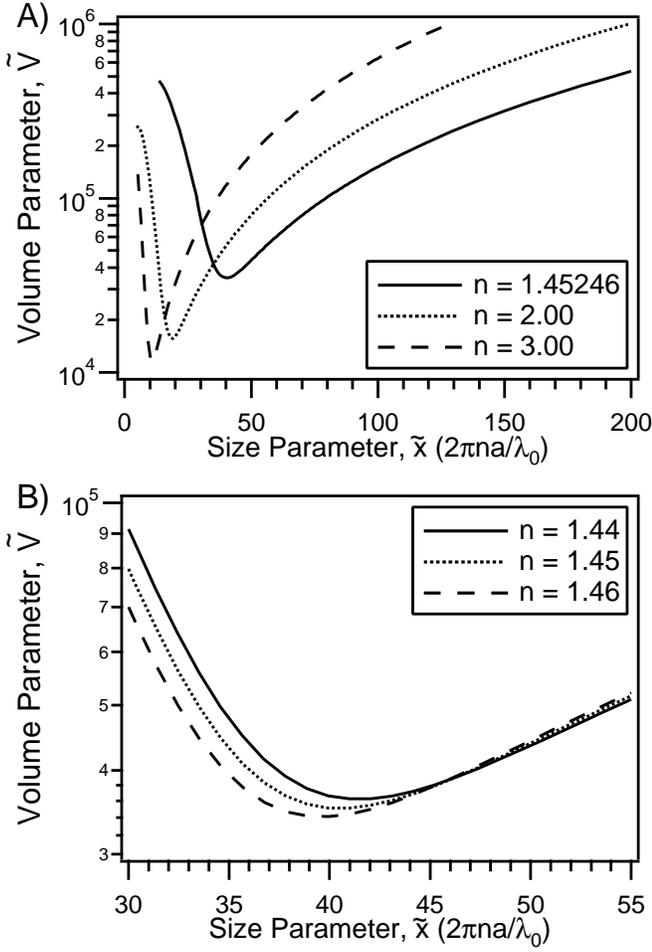}
\caption{\label{vol_param}(a) The dimensionless volume parameter,
$\tilde{V}$ (defined by Eq. \ref{volparam}), as a function of the
dimensionless size parameter, $\tilde{x}$ (defined by Eq.
\ref{size_param}). The solid line is for an index of refraction
$n=1.45246$, the index of refraction for fused silica at
$\lambda_{0} =852\mathrm{nm}$, with a minimum of $\tilde{V}
=34883.4$ for $\tilde{x} =39.9469$ $(l=m=34)$. The dotted line is
for an index of refraction $n=2.00$, with a minimum of $\tilde{V}
=15596.2$ for $\tilde{x} =18.9864$ $(l=m=14)$. The dashed line is
for an index of refraction $n=3.00$, with a minimum of $\tilde{V}
=11546.4$ for $\tilde{x} =10.2748$ $(l=m=6)$. (b) Because the
index of refraction for fused silica varies from $n=1.444$ at
$\lambda_{0}=1550\mathrm{nm}$ to $n=1.458$ for
$\lambda_{0}=600\mathrm{nm}$ (see Fig.~\ref{index}), this plot of
the dimensionless volume parameter, $\tilde{V}$, as a function of
the dimensionless size parameter, $\tilde{x}$, is made for that
range of values. The solid line is for an index of refraction
$n=1.44$, with a minimum of $\tilde{V} =36247.5$ for $\tilde{x}
=40.9812$, $(l=m=35)$. The dotted line is for an index of
refraction $n=1.45$, with a minimum of $\tilde{V} =35161.1$ for
$\tilde{x} =41.0036$, $(l=m=35)$. The dashed line is for an index
of refraction $n=1.46$, with a minimum of $\tilde{V} =34129.1$ for
$\tilde{x} =39.9631$, $(l=m=34)$.}
\end{figure}

In order to derive an answer that can be applied to different
wavelengths, one can define a dimensionless mode volume parameter,
$\tilde{V}$, and plot as a function of a dimensionless sphere size
parameter, $\tilde{x}$, defined as:
\begin{equation}
\tilde{V} = \frac{V_{\vec{P}}}{(\frac{\lambda_{0}}{2 \pi n})^3}
\label{volparam}
\end{equation}
and
\begin{equation}
\tilde{x} = \frac{2 \pi n a}{\lambda_{0}}, \label{size_param}
\end{equation}
where $V_{\vec{P}}$ is the cavity mode volume, $n$ is the index of
refraction at the free-space wavelength $\lambda_{0}$, and $a$ is
the sphere radius. The plots then only depend on the index of
refraction (see Fig.~\ref{vol_param}).

\begin{figure}[tb!]
  \includegraphics[width=8.6cm]{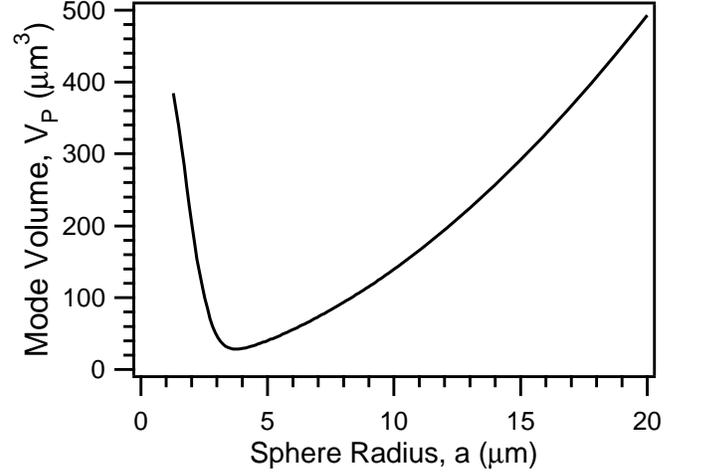}
  \caption{\label{vol_852}The electromagnetic mode volume, $V_{\vec{P}}$, for the TM
modes of a very low-OH fused silica microsphere as a function of
sphere radius at the wavelength $\lambda_{0} =852\mathrm{nm}$ for
the $D_{2}$ line of atomic Cesium. The minimum,
$28.4\mathrm{\protect\mu m}^{3}$, occurs for radius $a_{0}\approx
3.73\mathrm{\protect\mu m}$ corresponding to mode numbers $p=1$
and $l=m=34$.}
\end{figure}

Naively, one might assume that the sphere should be made as small
as possible in order to minimize the electromagnetic mode volume,
and hence to provide a maximum for $g_{0}$ and hence globally for
$g\left(\vec{r}\right)$. However, as shown in
Figs.~\ref{vol_param} and \ref{vol_852}, the mode volume for the
TM modes of a quartz microsphere actually passes through a minimum
at some particular radius $a_{0}$. This behavior can be understood
by noting that for $a<a_{0}$, the intrinsic, radiative losses are
increasing rapidly and ultimately cause the mode to no longer be
well-confined by the sphere, with a concomitant increase of the
mode volume. Note that in Fig.~\ref{vol_param} and subsequent
figures, we give results for $n\sim 1.45$ corresponding to fused
silica, as well as for $n=2.00$ and $n=3.00$. These latter cases
serve to illuminate the role of $n$ as well as being applicable to
other materials (i.e., the index of refraction for GaAs is $n=3.4$
for $\lambda=1550\mathrm{nm}$ \cite{h_optics}). For a very low-OH
fused silica microsphere at $\lambda_{0} =852\mathrm{nm}$ (the
wavelength of the $D_{2}$ transition in atomic Cesium) with index
of refraction $n=1.45246$, the minimum mode volume
$V_{\vec{P}}^{\min }\approx 28.4\mu \mathrm{m}^{3}$ occurs for
radius $a\approx 3.73\mu \mathrm{m}$ corresponding to mode numbers
$p=1,l=m=34$ (see Fig.~\ref{vol_852}). One might at first believe
that this value for the radius represents the optimal sphere size
for use as a cavity with single atoms. However, while the mode
volume $V_{\vec{P}}$ plays an important role in determining the
coupling constant (Eq. \ref{geezero}), it is not the only
parameter relevant to cavity QED with single atoms in a regime of
strong coupling. As discussed in the next sections, the quality
factor, $Q$, of a WGM has a strong dependence on the sphere
radius, and must also be considered in an attempt to optimize the
critical atom and saturation photon numbers.

\begin{figure}[tb!]
\includegraphics[width=8.6cm]{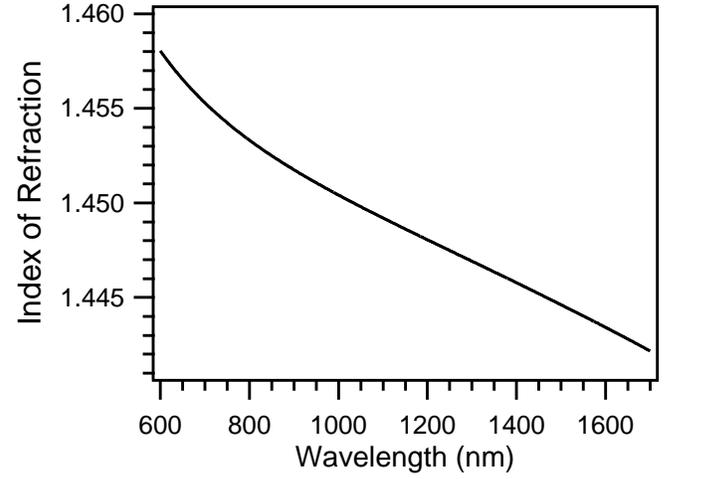}
\caption{\label{index}The index of refraction of very low-OH fused
silica as a function of wavelength.}
\end{figure}

\section{Losses in Dielectric Spheres}\label{losses}

\begin{figure}[tb!]
\includegraphics[width=8.6cm]{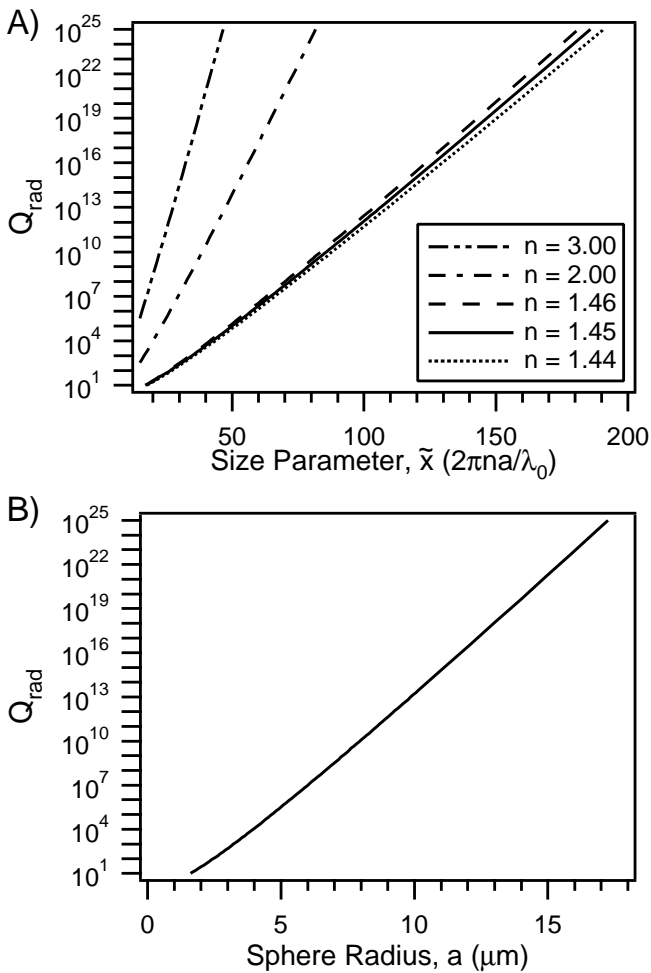}
\caption{\label{q_rad}(a) Semi-log plot of the radiative quality
factor, $Q_{rad}$, for various indices of refraction as a function
of the dimensionless size parameter, $\tilde{x} =\frac{2 \pi n
a}{\lambda_{0}}$. (b) Semi-log plot of the radiative quality
factor, $Q_{rad}$, as a function of sphere radius for a wavelength
of $\lambda_{0} = 852.359\mathrm{nm}$ (index of refraction is
$n=1.45246$).}
\end{figure}

For fused silica spheres with radius $a\gtrsim 15\mu\mathrm{m}$,
the effect of intrinsic radiative losses can be safely neglected,
since they allow quality factor $Q\gtrsim 10^{21}$, as illustrated
in Fig.~\ref{q_rad}. Such large values of $Q$ greatly exceed those
imposed by technical constraints of material properties, such as
bulk absorption and surface scattering.

However, as one moves to very small spheres with radius $a\lesssim
10\mu\mathrm{m}$, the intrinsic radiative $Q$ falls steeply enough
to become the dominant loss mechanism even in the face of other
technical imperfections. When assessing the usefulness of
microspheres for cavity QED, one must account for the entire set
of loss mechanisms to determine the optimal size for the
microsphere, which is the subject to which we now turn our
attention.

The quality factors of the WGMs of fused silica microspheres are
determined by several different loss mechanisms. The overall
quality factor can then be calculated by adding the different
contributions in the following way \cite{gorodetsky}:
\begin{eqnarray}
Q^{-1} &=& Q_{\mathrm{rad}}^{-1}+Q_{\mathrm{mat}}^{-1},  \label{q} \\
Q_{\mathrm{mat}}^{-1} &=&
Q_{\mathrm{s.s.}}^{-1}+Q_{\mathrm{w}}^{-1}+Q_{\mathrm{bulk}}^{-1},\label{qmat}
\end{eqnarray}
where $Q_{\mathrm{rad}}$ is due to purely radiative losses for an
ideal dielectric sphere and $Q_{\mathrm{mat}}$ results from
non-ideal material properties. The principal mechanisms
contributing to $Q_{\mathrm{mat}}$ are scattering losses from
residual surface inhomogeneities ($Q_{\mathrm{s.s.}}$), absorption
losses due to water on the surface of the sphere
($Q_{\mathrm{w}}$), and bulk absorption in the fused silica
($Q_{\mathrm{bulk}}$). The intrinsic material losses are known
very accurately, since they arise from absorption in the material
at the wavelength of concern \cite{lin}. Considerably greater
uncertainty is associated with the losses due to surface
scattering and absorption due to adsorbed material on the surface
of the sphere, of which water is likely the principal component.
We will adopt the models for these losses presented in
Refs.~\cite{gorodetsky,vernooy2}, extrapolated to the regime of
small spheres of interest here.

\subsection{Intrinsic Radiative Losses}

The contribution to the quality factor for purely radiative
effects, $Q_{\mathrm{rad}}$, can be derived by following the
arguments presented in Ref.~\cite{datsyuk}. These losses are due
to the leakage of light from the resonator due to its finite
dielectric constant and radius of curvature. The results can then
be compared to numerical results obtained by Lorenz-Mie
theory~\cite{lin2}. We find from Ref.~\cite{datsyuk} that
\begin{equation}
Q_{\mathrm{rad}}=\frac{1}{2}\left( l+\frac{1}{2}\right)
n^{1-2b}\left( n^{2}-1\right) ^{1/2}e^{2T_{l}},\label{qrad}
\end{equation}
where
\begin{equation}
T_{l}=\left( l+\frac{1}{2}\right) \left( \eta _{l}-\tanh \eta
_{l}\right) ,
\end{equation}
\begin{equation}
\eta _{l}=\mathrm{arccosh}\left\{ n\left[
1-\frac{1}{l+\frac{1}{2}}\left( t_{p}^{0}\xi
+\frac{l^{1-2b}}{\sqrt{l^{2}-1}}\right) \right] ^{-1}\right\},
\end{equation}
\begin{equation}
\xi =\left[ \frac{1}{2}\left( l+\frac{1}{2}\right) \right]
^{\frac{1}{3}},
\end{equation}
and
\begin{equation} b=\begin{cases} 0& \text{TE modes},\\ 1& \text{TM
modes}.\end{cases}
\end{equation}
Also, $n$ is the index of refraction and $t_{p}^{0}$ is the
$p^{th}$ zero of the Airy function $Ai$. This $p$ corresponds to
the mode number $(p,l,m)$. In our case, we are only interested in
the $p=1$ modes of the sphere to maximize the electromagnetic
field outside the sphere while maintaining a small mode volume.
Note that these expressions for $Q_{rad}$ become invalid in the
limit of small $l$ mode numbers. The error in the mode functions
used to derive these results reaches $1\%$ for $l=18$. However,
the error is less than $0.2\%$ for $l=76$ (This is the optimal
sphere size discussed in Section~\ref{strong_coupling_cesium}).
Fortunately, the expressions are valid in the regimes for which we
are concerned. This has been confirmed by making comparisons with
numerical values obtained using Lorenz-Mie scattering theory.

From Fig.~\ref{q_rad}, we see that the radiative Q falls
approximately exponentially as the radius $a$ is decreased, and
can become quite important as the sphere size is decreased below
$10\mathrm{\mu m}$. For example, for a $15\mathrm{\mu m}$ radius
sphere and a wavelength $\lambda_{0} = 852.359\mathrm{nm}$,
$Q_{\mathrm{rad}}\approx 2\times 10^{21}$. Therefore, the net
quality factor would most certainly be dominated by other loss
mechanisms in Eq.~\ref{q}. However, for a $7\mathrm{\mu m}$ radius
sphere, $Q_{\mathrm{rad}}\approx 4\times 10^{8}$, and the
radiative losses can play a crucial role in the characteristics of
the spheres that are optimal for use in cavity QED.

\subsection{Material Loss Mechanisms}

\begin{figure}[tb]
\includegraphics[width=8.6cm]{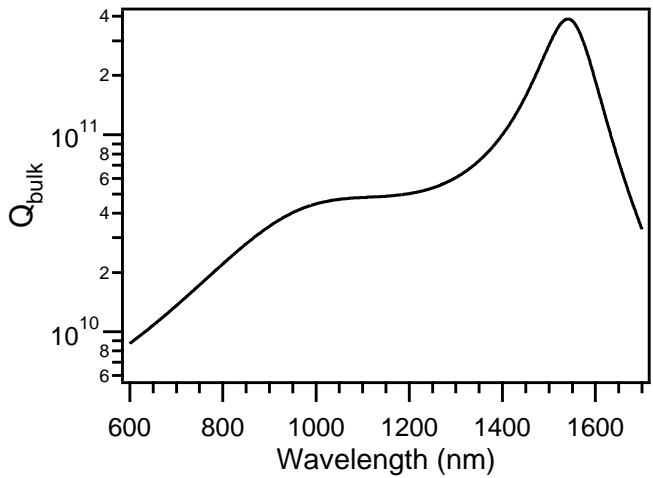}
\caption{\label{qbulk}The quality factor, $Q_{\mathrm{bulk}}$,
from Eq. \protect\ref{qbulkeq} for a very low-OH fused silica
microsphere as a function of wavelength. Because fused silica has
a minimum in absorption at $1550 \mathrm{nm}$, there is a maximum
for the quality factor due to bulk absorption of
$Q_{\mathrm{bulk}}\sim 3.8\times 10^{11}$. At $852 \mathrm{nm}$,
the quality factor due to bulk absorption is
$Q_{\mathrm{bulk}}\sim 2.4\times 10^{10}$.}
\end{figure}

The quality factor due to bulk absorption, $Q_{\mathrm{bulk}}$, in
fused silica is actually known very well, since this depends only
on the absorption of the material at the wavelength of concern
\cite{gorodetsky}:
\begin{equation}
Q_{\mathrm{bulk}}=\frac{2\pi n}{\alpha \lambda_{0}},
\label{qbulkeq}
\end{equation} where $n$ is the index of refraction, and
$\alpha$ is the absorption coefficient of the material. From
Fig.~\ref{qbulk} we see that for very low-OH fused silica, the
absorption coefficient at $852\mathrm{nm}$ is $\alpha\approx
4.5\times 10^{-4}\mathrm{m}^{-1}$ \cite{lin}. This would
correspond to a quality factor of $Q_{\mathrm{bulk}}\sim 2.4\times
10^{10}$. Fused silica has a minimum in its absorption coefficient
of $\alpha\approx 1.5\times 10^{-5}\mathrm{m}^{-1}$ at $1550
\mathrm{nm}$, which yields a quality factor of
$Q_{\mathrm{bulk}}\sim 3.8\times 10^{11}$.

The quality factor due to surface scattering, $Q_{\mathrm{s.s.}}$,
and absorption by adsorbed water, $Q_{\mathrm{w}}$, has also been
studied and modelled, albeit for larger spheres with $a\gtrsim
600\mu\mathrm{m}$. For losses due to surface scattering, we follow
the work of Refs.~\cite{gorodetsky,vernooy2} and take
\begin{equation}
Q_{\mathrm{s.s.}}\sim \frac{3\varepsilon (\varepsilon
+2)^{2}}{(4\pi)^{3}(\varepsilon -1)^{5/2}}\frac{\lambda_{0}
^{7/2}(2 a )^{1/2}}{(\sigma B)^{2}},
\end{equation}
where $\varepsilon =n^{2}$ is the dielectric constant and $\sigma
B\sim 5\mathrm{nm}^{2}$ is an empirical parameter determined by
the size and correlation length of the distribution of residual
surface inhomogeneities. This quantity was reported in
Ref.~\cite{vernooy2} based upon atomic force microscopy
measurements of a microsphere.

The quality factor due to water adsorbed on the surface,
$Q_{\mathrm{w}}$, is given by \cite{vernooy2}
\begin{equation}
Q_{\mathrm{w}}\sim \sqrt{\frac{\pi }{8n^{3}}}\frac{(2 a
)^{1/2}}{\delta \lambda_{0}^{1/2}\beta _{w}},
\end{equation}
where $\delta \sim 0.2\mathrm{nm}$ is an estimated thickness for
the water layer, and $\beta _{w}\sim 4.33\mathrm{m}^{-1}$ is the
absorption coefficient of water at $852\mathrm{nm}$.

\begin{figure}[tb]
\includegraphics[width=8.6cm]{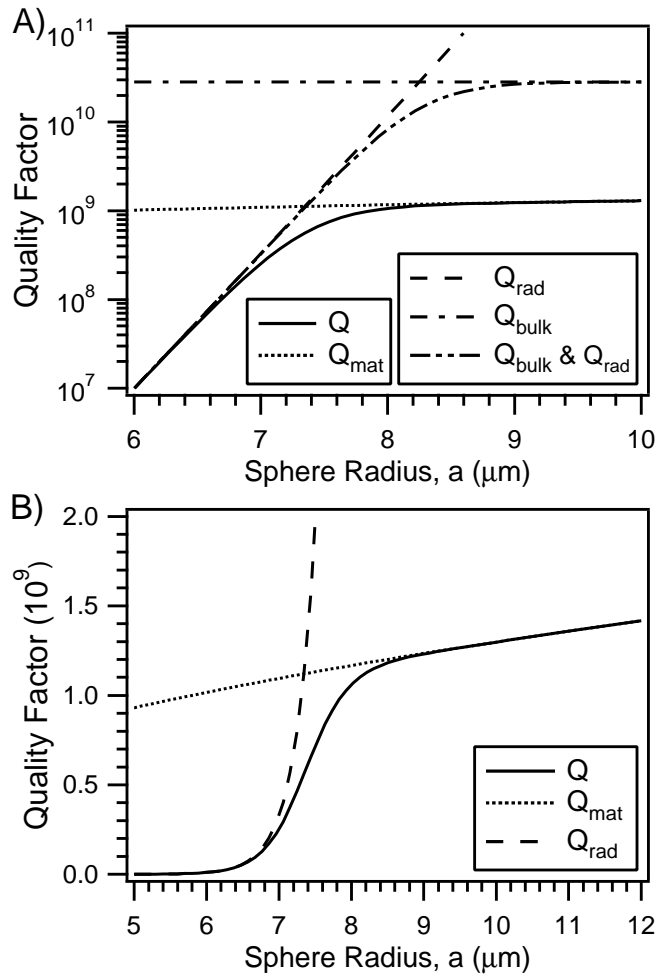}
\caption{\label{q_plot_852}(a) Semi-log plot of the quality
factors due to the various loss mechanisms discussed in Section
\ref{losses} for a very low-OH fused silica microsphere as a
function of sphere radius for the $l=m$, TM modes at a wavelength
of $\protect\lambda_{0} =852\mathrm{nm}$. In particular, traces
are shown for the quality factor due to purely radiative losses
($Q_{\mathrm{rad}}$), the bulk absorption of fused silica
($Q_{\mathrm{bulk}}$), both radiative losses and bulk absorption,
the three loss mechanisms comprising $Q_{\mathrm{mat}}$:
($Q_{\mathrm{bulk}}$, $Q_{\mathrm{s.s.}}$, $Q_{\mathrm{w}}$), and
the predicted $Q$ due to all four loss mechanisms. (b) This linear
plot zooms in on the region of interest at the transition where
the radiative losses become the dominant loss mechanism. The plot
contains the quality factor due to purely radiative losses
($Q_{\mathrm{rad}}$), the three loss mechanisms comprising
$Q_{\mathrm{mat}}$: ($Q_{\mathrm{bulk}}$, $Q_{\mathrm{s.s.}}$,
$Q_{\mathrm{w}}$), and the predicted $Q$ due to all four loss
mechanisms.}
\end{figure}

Combining these various results, we display in
Fig.~\ref{q_plot_852} a curve for the quantity $Q_{\mathrm{mat}}$
as a function of sphere radius, $a$, for a wavelength $\lambda_{0}
= 852\mathrm{nm}$. This same figure shows the quality factor,
$Q_{\mathrm{rad}}$, set by intrinsic radiative losses (Eq.
\ref{qrad}), as well as the overall quality factor,
$Q=\frac{Q_{\mathrm{rad}}Q_{\mathrm{mat}}}{Q_{\mathrm{rad}}+Q_{\mathrm{mat}}}$.
From this plot, we see that the radiative losses dominate the
overall quality factor below a radius of $a\lesssim 8\mathrm{\mu
m}$, while the losses due to material properties are most
significant for $a\gtrsim 8\mathrm{\mu m}$. Because of the
extremely steep dependence of $Q_{\mathrm{rad}}$ on sphere size,
the point of transition from material to radiative dominated loss
should be reasonably insensitive to details of the models employed
to describe the material losses. Although we focus our attention
here on the wavelength appropriate to the particular case of the
$D_{2}$ transition in atomic Cesium, a similar analysis could be
carried out for other wavelengths of interest using the above
formalism, as for example the $2S\rightarrow 2P$ transition at
$1.083\mathrm{\mu m}$ in metastable Helium.

\section{The Strong Coupling Regime}\label{strong_coupling}

The ultimate goal that we consider here is to employ the WGMs of
quartz microspheres as cavity modes for achieving strong coupling
to atoms within the setting of cavity QED. The atom of choice in
this paper is Cesium, and in particular, the $D_2$ ($F=4\mapsto
F^{\prime}=5$) transition in Cesium at $\lambda_{0}
=852.359\mathrm{nm}$ as an illustrative example. Such an analysis
allows a direct comparison with the state of the art in
Fabry-Perot cavities \cite{hoodye}.

The coupling coefficient $g(\vec{r})$ is the coupling frequency of
a single atom to a particular cavity mode and corresponds to
one-half the single photon Rabi frequency \cite{jeff,jeff2}. For
an atom located just at the outer surface of the microsphere
(i.e., in vacuum) and interacting with a whispering gallery mode
$\vec{P}=(p,l,m)$, the coupling coefficient is given by
\cite{vernooy1}
\begin{equation}
g(a)\equiv g_{a}=\gamma _{\bot}\left\vert
\vec{\Psi}_{\mathrm{out}}(a)\right\vert
\sqrt{\frac{V_{0}}{V_{\vec{P}}}},
\end{equation}
where $a$ is the sphere radius,
$\frac{\gamma_{\bot}}{2\pi}=2.61\mathrm{MHz}$ is the transverse
spontaneous decay rate for our transition in Cesium,
$V_{0}=\frac{3c\lambda_{0}^{2}}{4\pi \gamma _{\bot}}$ is the
effective volume of the atom for purely radiative interactions,
and $V_{\vec{P}}$ is the electromagnetic mode volume of the
whispering gallery mode designated by $\vec{P}=(p,l,m)$.

Armed with a knowledge of $g$, we are now able to determine
certain dimensionless parameters relevant to the strong coupling
regime of cavity QED. In particular, we consider an atom-cavity
system to be in the strong coupling regime when the single-photon
Rabi frequency, $2g$, for a single intracavity atom dominates the
cavity field decay rate, $\kappa $, the atomic dipole decay rate,
$\gamma_{\bot}$, and the inverse atomic transit time, $T^{-1}$
\cite{jeff,jeff2}. We will defer further discussion of $T^{-1}$,
however, this requirement relates to the need for atomic
localization~\cite{vernooy1, mabuchi}. In the strong coupling
regime, important parameters for characterizing the atom-cavity
system are the two dimensionless parameters: the saturation photon
number, $n_{0}$, and the critical atom number, $N_{0}$. The
saturation photon number, given by
\begin{equation}
n_{0}\equiv \frac{\gamma_{\bot}^{2}}{2g^{2}},
\end{equation}
corresponds to the number of photons required to saturate an
intracavity atom \cite{jeff,jeff2}. The critical atom number,
defined by
\begin{equation}
N_{0}\equiv \frac{2\gamma_{\bot}\kappa}{g^{2}},
\end{equation}
corresponds to the number of atoms required to have an appreciable
effect on the intracavity field \cite{jeff,jeff2}. Ideally, one
hopes to minimize simultaneously both the critical atom number,
$N_{0}$, and the saturation photon number, $n_{0}$, which
corresponds to simultaneous maxima for both $\frac{g^{2}}{\kappa
\gamma_{\bot}}$ and $\frac{g^2}{\gamma_{\bot}^{2}}$.

The saturation photon number and critical atom number are useful
because of their physical meaning. However, one can define a new
dimensionless parameter
\begin{equation}
\beta = \frac{8\pi^2 V_{\vec{P}}}{3
\lambda_{0}^{3}}\frac{1}{\left\vert
\vec{\Psi}_{\mathrm{out}}(a)\right\vert ^{2}}, \label{beta}
\end{equation}
that corresponds to the cavity mode volume in units of $\lambda^3$
weighted by the inverse of the strength of the mode function at
the atomic position. This enables the equations for the saturation
photon number and critical atom number to be expressed as:
\begin{equation}
n_{0} = \frac{\beta}{4 Q_{\mathrm{atom}}}, \label{satbeta}
\end{equation}
and
\begin{equation}
N_{0} = \frac{\beta}{Q_{\mathrm{cavity}}}, \label{critbeta}
\end{equation}
where
\begin{equation}
Q_{\mathrm{atom}} = \frac{\pi c}{\lambda_{0} \gamma_{\bot}},
\label{qatom}
\end{equation}
and
\begin{equation}
Q_{\mathrm{cavity}} = \frac{\pi c}{\lambda_{0} \kappa}.
\label{qcavity}
\end{equation}
This parameter, $\beta$, then also determines the coupling
coefficient in the following manner:
\begin{equation}
g(a)=\sqrt{\frac{2\pi c \gamma_{\bot}}{\beta \lambda_{0}}}.
\end{equation}
Therefore, we see that one can use a single parameter, $\beta$,
combined with the properties of the atom to be used ($\lambda_{0}$
and $\gamma_{\bot}$) and the quality factor of the resonator,
$Q_{\mathrm{cavity}}$, to determine the three parameters
($n_{0},N_{0},g_{0}$) of importance in determining the quality of
an atom-cavity system.

\begin{figure}[tb]
\includegraphics[width=8.6cm]{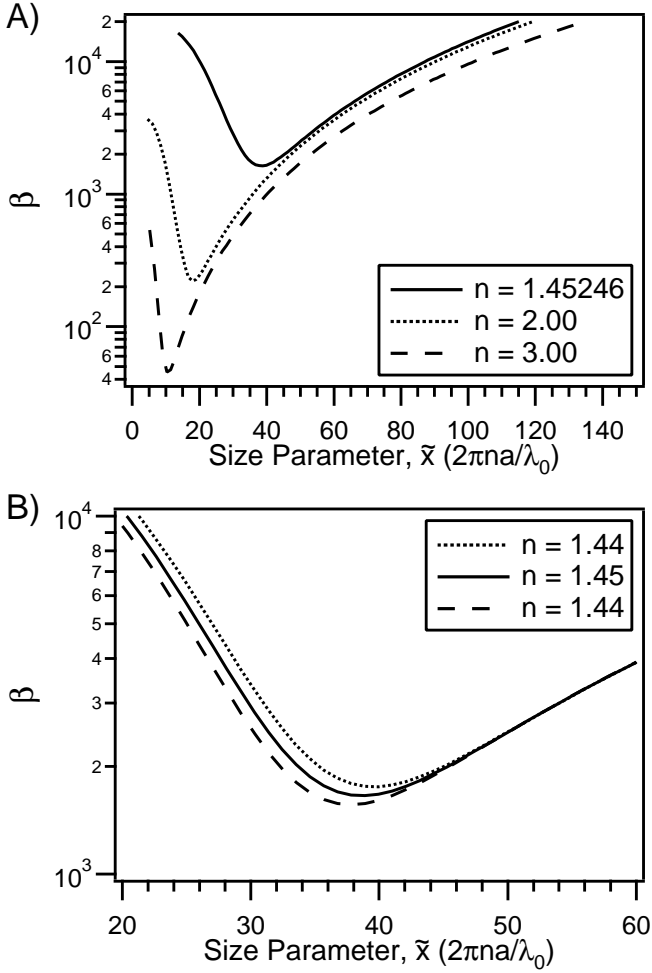}
\caption{\label{beta_plot} (a) The dimensionless parameter $\beta$
as a function of the dimensionless size parameter $\tilde{x}
=\frac{2 \pi n a}{\lambda_{0}}$. For an index of refraction
$n=1.45246$ (i.e., the index of refraction for fused silica at
$\lambda_{0} =852\mathrm{nm}$), there is a minimum of $\beta
=1632.01$ for $\tilde{x} =38.8833$, $(l=m=33)$. For an index of
refraction $n=2.00$, there is a minimum of $\beta =221.124$ for
$\tilde{x} =17.8763$, $(l=m=13)$. For an index of refraction
$n=3.00$, there is a minimum of $\beta =45.3744$ for $\tilde{x}
=10.2748$, $(l=m=6)$. (b) Because the index of refraction for
fused silica varies from $n=1.444$ at
$\lambda_{0}=1550\mathrm{nm}$ to $n=1.458$ for
$\lambda_{0}=600\mathrm{nm}$ (see Fig.~\ref{index}), this plot is
made for that range of values. For an index of refraction
$n=1.44$, there is a minimum of $\beta =1753.92$ for $\tilde{x}
=39.9188$, $(l=m=34)$. For an index of refraction $n=1.45$, there
is a minimum of $\beta =1653.7$ for $\tilde{x} =38.8778$,
$(l=m=33)$. For an index of refraction $n=1.46$, there is a
minimum of $\beta =1561.45$ for $\tilde{x} =37.8348$, $(l=m=32)$.}
\end{figure}

\begin{figure}[tb]
\includegraphics[width=8.6cm]{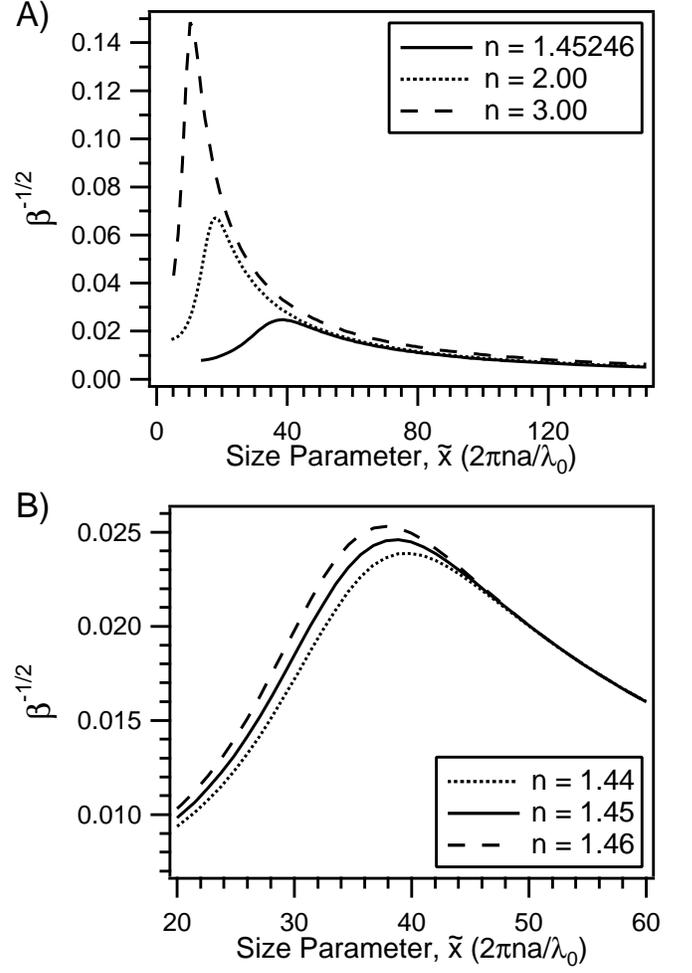}
\caption{\label{invsqrt_b} (a) The dimensionless parameter
$\frac{1}{\sqrt{\beta}}$ as a function of the dimensionless size
parameter $\tilde{x} =\frac{2 \pi n a}{\lambda_{0}}$. For an index
of refraction $n=1.45246$ (i.e., the index of refraction for fused
silica at $\lambda_{0} =852\mathrm{nm}$), there is a maximum of
$\frac{1}{\sqrt{\beta}} =0.0247536$ for $\tilde{x} =38.8833$,
$(l=m=33)$. For an index of refraction $n=2.00$, there is a
maximum of $\frac{1}{\sqrt{\beta}} =0.0672484$ for $\tilde{x}
=17.8763$, $(l=m=13)$. For an index of refraction $n=3.00$, there
is a maximum of $\frac{1}{\sqrt{\beta}} =0.148455$ for $\tilde{x}
=10.2748$, $(l=m=6)$. (b) Because the index of refraction for
fused silica varies from $n=1.444$ at
$\lambda_{0}=1550\mathrm{nm}$ to $n=1.458$ for
$\lambda_{0}=600\mathrm{nm}$ (see Fig.~\ref{index}), this plot is
made for that range of values. For an index of refraction
$n=1.44$, there is a maximum of $\frac{1}{\sqrt{\beta}}
=0.0238779$ for $\tilde{x} =39.9188$, $(l=m=34)$. For an index of
refraction $n=1.45$, there is a minimum of $\frac{1}{\sqrt{\beta}}
=0.0245908$ for $\tilde{x} =38.8778$, $(l=m=33)$. For an index of
refraction $n=1.46$, there is a minimum of $\frac{1}{\sqrt{\beta}}
=0.0253068$ for $\tilde{x} =37.8348$, $(l=m=32)$.}
\end{figure}

Figs.~\ref{beta_plot} and \ref{invsqrt_b} are plots of this
dimensionless parameter $\beta$ and of $\frac{1}{\sqrt{\beta}}$ as
functions of the dimensionless size parameter $\tilde{x} =\frac{2
\pi n a}{\lambda_{0}}$ for a few values of index of refraction.
Because the index of refraction for fused silica varies from
$n=1.444$ at $\lambda_{0}=1550\mathrm{nm}$ to $n=1.458$ for
$\lambda_{0}=600\mathrm{nm}$ (see Fig.~\ref{index}),
Figs.~\ref{beta_plot}b and \ref{invsqrt_b}b are made for that
range of values. From Figs.~\ref{beta_plot} and \ref{invsqrt_b}
one sees that there is a minimum for $\beta$ and a maximum for
$\frac{1}{\sqrt{\beta}}$ that depends on the index of refraction.

\section{Strong Coupling with Cesium}\label{strong_coupling_cesium}

\begin{figure}[tb]
\includegraphics[width=8.6cm]{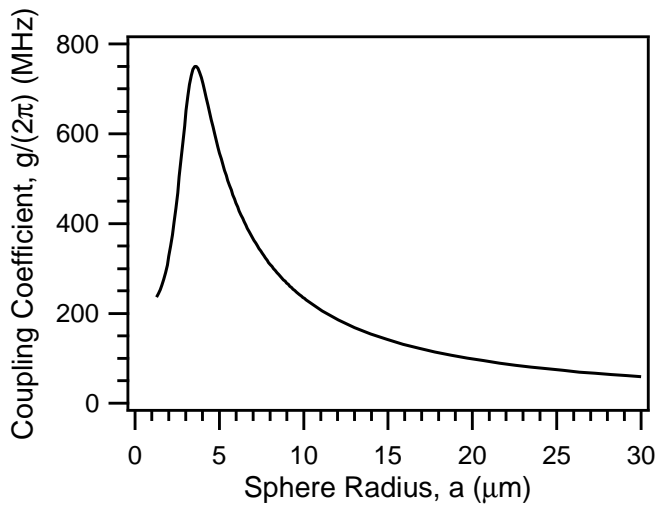}
\caption{\label{g_852} The coupling coefficient, $\frac{g}{2\pi
}$, as a function of sphere size for the $D_{2}$ transition in
Cesium ($\lambda_{0} =852.359 \mathrm{nm}$). There is a maximum of
$\frac{g}{2\pi } = 749.986 \mathrm{MHz}$ for a sphere radius of
$a=3.63163\mu\mathrm{m}$, ($l=m=33$). Note that the maximum for
$\frac{g}{2\pi }$ does not coincide with the minimum for the
cavity mode volume, $V_{\vec{P}}$ (see Fig.~\ref{vol_852}).}
\end{figure}

The results of the previous section can now be used to determine
the saturation photon number, $n_{0}$, the critical atom number,
$N_{0}$, and the coupling coefficient, $g(a)$, for any atomic
transition. In our case, we are concerned with the $D_{2}$
transition in Cesium ($\lambda_{0} =852.359\mathrm{nm}$). For this
transition, the spontaneous transverse decay rate is
$\frac{\gamma}{2 \pi} = 2.61 \mathrm{MHz}$. Also, at this
wavelength the index of refraction for fused silica is $n =
1.45246$. This allows one to compute the coupling coefficient,
$g(a)=\sqrt{\frac{2\pi c \gamma_{\bot}}{\beta \lambda_{0}}}$.
Fig.~\ref{g_852} shows that there is a maximum of $\frac{g}{2 \pi}
= 749.986 \mathrm{MHz}$ for a radius $a=3.63\mu\mathrm{m}$,
($l=m=33$). Interestingly, because we are restricted to having the
atom couple to the \textit{external} field of the microsphere, the
maximum in the coupling coefficient, $g(a)$, does not coincide
with the minimum for the mode volume, $V_{\vec{P}}$ (see
Figs.~\ref{vol_852} and \ref{g_852}.)

\begin{figure}[tb]
\includegraphics[width=8.6cm]{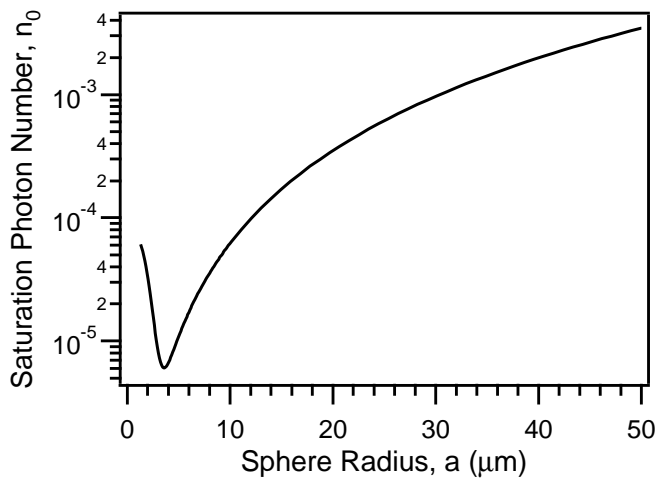}
\caption{\label{sat_n852}The saturation photon number, $n_{0}$, as
a function of sphere size for the $D_2$ transition in Cesium
($\lambda_{0} = 852.359\mathrm{MHz}$). There is a minimum $n_{0}=
6.05527\times 10^{-6}$ for a sphere radius of $a
=3.63163\mu\mathrm{m}$ $(l=m=33)$. At this radius, the coupling
coefficient is $\frac{g}{2\pi}=749.986 \mathrm{MHz}$. }
\end{figure}

The saturation photon number, $n_{0}$, is proportional to the
dimensionless parameter $\beta$ as shown in Eq. \ref{satbeta}.
Since the factor of proportionality is a constant that depends
only on the properties of the particular atom of concern, the
curve is determined by that of $\beta$ along with the quality
factor of the atomic resonance (in our case Cesium), which is
given by Eq. \ref{qatom} to be $Q_{atom} = 6.738\times 10^{7}$.
Fig.~\ref{sat_n852} is a plot of the saturation photon number for
the $D_2$ transition in Cesium as a function of sphere size.
Fig.~\ref{sat_n852} shows that there is a minimum for the
saturation photon number of $n_{0}= 6.05527\times 10^{-6}$ for a
sphere radius of $a =3.63163\mu\mathrm{m}$ $(l=m=33)$.

\begin{figure}[tb]
\includegraphics[width=8.6cm]{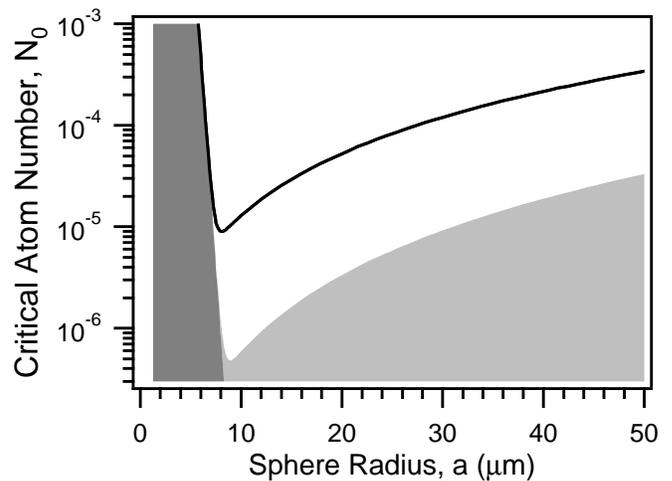}
\caption{\label{crit_n852}The critical atom number, $N_{0}$, as a
function of sphere size for the $D_2$ transition in Cesium
($\lambda_{0} = 852.359\mathrm{MHz}$). There is a minimum $N_{0}=
8.99935\times 10^{-6}$ for a sphere radius of $a
=8.12015\mu\mathrm{m}$ $(l=m=79)$. At this radius, the coupling
coefficient is $\frac{g}{2\pi}=304.16 \mathrm{MHz}$. This plot of
the critical atom number incorporates the model for the quality
factor of the resonator, $Q_{cavity}$, outlined in section
\ref{losses}, for the four loss mechanisms: bulk absorption,
surface scattering, absorption due to water on the surface, and
radiative losses. The dark grey region is bounded by the effects
of purely radiative losses. The light grey region is bounded by
the effects of both radiative losses and bulk absorption.}
\end{figure}

The critical atom number, $N_{0}$, is also proportional to the
dimensionless parameter $\beta$ as shown in Eq. \ref{critbeta}.
However, its factor of proportionality is the quality factor of
the resonator, $Q_{cavity}$, which has a very strong dependence on
the sphere radius, $a$, in the region below 10$\mathrm{\mu m}$
(see Fig.~\ref{q_plot_852}). Therefore, the minimum for the
critical atom number does not occur for the same sphere size as
for the saturation photon number. Fig.~\ref{crit_n852} is a plot
of the critical atom number as a function of sphere size. Using
for $Q_{cavity}$ the model that incorporates all of the loss
mechanisms discussed in section \ref{losses} (radiative losses,
bulk absorption, surface scattering, and absorption due to water
on the surface), we find that the minimum for the critical atom
number $N_{0}= 8.99935\times 10^{-6}$ occurs for a sphere radius
of $a =8.12015\mu\mathrm{m}$ $(l=m=79)$. At this radius, the
coupling coefficient is $\frac{g}{2\pi}=304.16 \mathrm{MHz}$.

\begin{figure}[tb]
\includegraphics[width=8.6cm]{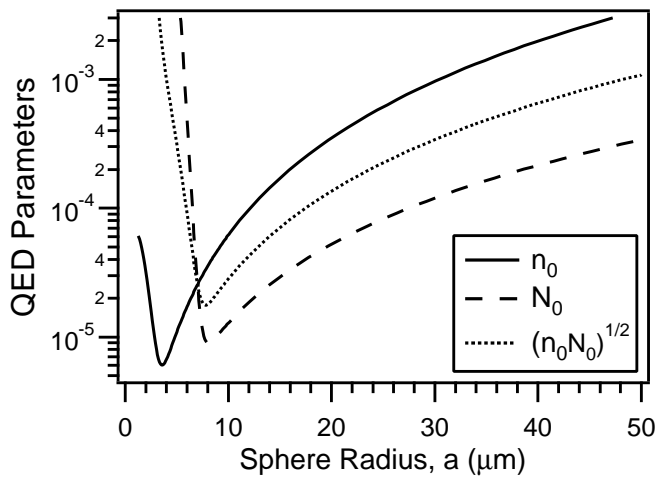}
\caption{\label{qedparam_n852} This plot shows the two parameters,
$(n_{0},N_{0})$, of importance to cavity QED as a function of
sphere radius. The geometric mean of these two parameters is also
plotted. The solid line represents the saturation photon number,
$n_{0}$, the dashed line gives the critical atom number, $N_{0}$,
and the dotted line shows the geometric mean of the two
parameters, $\sqrt{n_{0}\times N_{0}}$. The minimum of each plot
corresponds to the following dimensionless parameters:
$n_{0}=6.05527\times 10^{-6}$ for $a=3.63163\mathrm{\protect\mu
m}$ ($l=m=33$), and $N_{0}=8.99935\times 10^{-6}$ at
$a=8.12015\mathrm{\protect\mu m}$ ($l=m=79$). The two curves cross
at $a=7.03\mathrm{\protect\mu m}$ with $n_{0}=N_{0}=2.56\times
10^{-5}$. The geometric mean of these two parameters,
$\sqrt{n_{0}\times N_{0}}$, is minimized for
$a=7.83038\mathrm{\protect\mu m}$ ($l=m=76$). For this radius, the
parameters are: $n_{0}=3.36107\times 10^{-5}$ and
$N_{0}=9.27834\times 10^{-6}$. Note that the curve for $N_{0}$
assumes the model for the $Q$ discussed in this paper, and that
the coupling coefficient $g\left(\vec{r}\right)$ is evaluated at
the maximum of the mode function for $r=a$.}
\end{figure}

Unfortunately, as illustrated in Fig.~\ref{qedparam_n852}, the
minima for the two parameters, $n_{0}$ and $N_{0}$, do not occur
for the same sphere radius. However, if one uses the minimum of
the geometric mean of the two parameters, each can have a value
near its respective minimum. The minimum of the geometric mean
occurs for a sphere radius $a=7.83038\mu\mathrm{m}$ $(l=m=76)$.
For this sphere size, the coupling coefficient is
$\frac{g}{2\pi}=318.333 \mathrm{MHz}$, the saturation photon
number is $n_{0}= 3.36107\times 10^{-5}$, and the critical atom
number is $N_{0}= 9.27834\times 10^{-6}$. Therefore, each cavity
QED parameter can be made to achieve simultaneously a value near
its respective minimum.

\begin{figure}[tb]
\includegraphics[width=8.6cm]{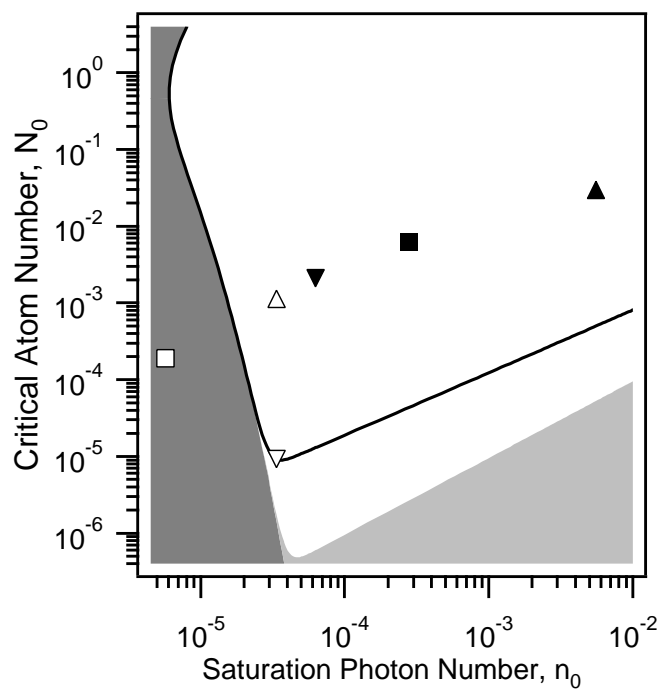}
\caption{\label{fabry_micro} The solid line gives a parametric
plot of the critical atom number, $N_{0}$, and the saturation
photon number, $n_{0}$, for fused-silica microspheres and the
$D_{2}$ transition of atomic Cesium, incorporating the loss
mechanisms outlined in section \ref{losses}. The dark grey region
is bounded by the effects of radiative losses. The light grey
region is bounded by the effects of bulk absorption and radiative
losses. This plot also offers a comparison of experimental and
theoretical cavity QED parameters for microsphere and Fabry-Perot
cavities. $\blacksquare $ represents the current state of the art
for cavity QED in Fabry-Perot cavities as in Ref.~\cite{science}.
$\square $ is a projection of the practical limit for Fabry-Perot
cavities based upon Ref.~\cite{hoodye}. $\blacktriangle $
represents the $60\protect\mu \mathrm{m}$ radius sphere
implemented for cavity QED in Ref.~\cite{vernooy3}.
$\blacktriangledown $ is the current state of the art in
$10\protect\mu\mathrm{m}$ microspheres based upon the results
presented in section \ref{strong_coupling_cesium}. $\vartriangle $
is the currently achievable $Q$ with the optimal sphere size of
$7.83\protect\mu \mathrm{m}$ based upon the analysis of sections
\ref{losses} and \ref{strong_coupling}. $\triangledown $ is the
theoretically achievable $Q\sim 9.76\times 10^{8}$ at the optimal
sphere size, $a\approx 7.83\protect\mu \mathrm{m}$.}
\end{figure}

\section{Progress in Small Sphere Manufacture}

A large portion of the work being done on microspheres has been to
push the quality factors of the spheres to record levels
\cite{gorodetsky,vernooy2}. This effort has produced some of the
highest finesse ($\mathcal{F}=2.3\times 10^{6}$) optical cavities
to date with quality factors $Q\sim 10^{10}$
\cite{gorodetsky,vernooy2}. However, we have seen that $Q$ is not
the only relevant factor in determining the suitability of the
WGMs for cavity QED in a regime of strong coupling. In general,
the preceding analysis demonstrates the requirement to push to
microspheres of small radius, $a\lesssim 10\mathrm{\mu m}$.
Unfortunately, the experiments that have achieved the highest
quality factors and which have investigated certain material loss
mechanisms are of rather larger size, and hence not optimal for
cavity QED in a regime of strong coupling. For example, the
experiment of Ref.~\cite{vernooy2} achieved a quality factor of
$Q=7.2\times 10^{9}$ at $850\mathrm{nm}$ in a sphere of radius
$a=340\mathrm{\mu m}$.

To explore the possibilities of cavity QED with strong coupling in
substantially smaller spheres, we have undertaken a program to
study fabrication techniques for quartz microspheres with
$a\lesssim 30\mathrm{\mu m}$, while still maintaining high quality
factors. We have been able to fabricate $10\mathrm{\mu m}$ radius
spheres using an oxygen-hydrogen micro-torch to melt the ends of
very low-OH fused silica rods to form a sphere on the end of a
stem. Light is then coupled to the sphere using frustrated total
internal reflection of a prism, as in Refs.~\cite{vernooy1,
vernooy2, vernooy3}. Our observations demonstrate that spheres of
this size can be made consistently to have quality factors
$Q\gtrsim 0.8\times 10^{7}$. While this is encouraging progress,
the resulting $Q$ is two orders of magnitude smaller than the
theoretical maximum of approximately $1.3\times 10^{9}$ for this
size based upon the model discussed in Section \ref{losses}.

One possible reason for this discrepancy could be the importance
of minimizing the ellipticity of the small spheres. Because the
small resonators fabricated by our technique have a stem
protruding out of them, they are far from spherical. When coupling
to an $l=m$ mode in spheres with $a\gtrsim 100\mathrm{\mu m}$ and
hence large $l$, the mode is tightly confined to the equator;
therefore, the poles do not have an appreciable impact on the mode
structure or quality factor. In this case, it is not of critical
importance to have the best sphere possible, but rather the best
great circle possible to achieve large quality factors. However,
this is not the case in small spheres with $a\lesssim
10\mathrm{\mu m}$. As $a$ decreases, the $l=m$ modes occupy an
increasingly larger proportion of the sphere in polar angle, and
the ellipticity of the sphere becomes increasingly important in
determining the mode structure as well as the $Q$. However, while
there is certainly room for improvement in our fabrication
technique and in the resulting mode structures and quality
factors, we shall see in the next section that the current results
have promising implications.

\section{Comparing Microspheres and Fabry-Perot Cavities}

Fig.~\ref{fabry_micro} offers a comparison of the state of the art
for Fabry-Perot and microsphere cavities for cavity QED, as well
as projections of likely limits for each. It is interesting to
note that in our projections for the limiting cases of each,
microspheres allow for a significant improvement in the critical
atom number, $N_{0}$, relative to Fabry-Perot cavities. On the
other hand, a principal advantage of Fabry-Perot cavities relative
to microspheres would seem to be significant improvements in the
saturation photon number, $n_{0}$. The specific specific task at
hand would then dictate which technology to apply.

As shown in Fig.~\ref{fabry_micro}, there has already been some
progress in coupling atoms to the external fields of a microsphere
\cite{vernooy3}. The sphere employed for the work of
Ref.~\cite{vernooy3} had a radius of $a\approx 60\mu \mathrm{m}$,
and quality factor $Q\lesssim 5\times 10^{7}$, corresponding to a
mode volume of $V_{\vec{P}}\approx 3.7\times 10^{3}\mu
\mathrm{m}^{3}$, coupling coefficient $g_{a}/(2\pi )\approx
24\mathrm{MHz}$, saturation photon number $n_{0}=5.54\times
10^{-3}$, and critical atom number $N_{0}=2.99\times 10^{-2}$. If
instead this experiment were to be implemented with a smaller
sphere with $10\mu \mathrm{m}$ radius and with quality factor
$Q\sim 0.8\times 10^{7}$ such as we have manufactured and
described in Section \ref{strong_coupling_cesium}, the following
parameters would be achieved: a mode volume of $V_{\vec{P}}\approx
1.4\times 10^{2}\mu\mathrm{m}^{3}$, coupling coefficient
$g_{a}/(2\pi )\approx 233\mathrm{MHz}$, saturation photon number
$n_{0}\approx 6.27\times 10^{-5}$, and critical atom number
$N_{0}\approx 2.11\times 10^{-3}$. Therefore, we see that
currently achievable quality factors in spheres of radius
$10\mathrm{\mu m}$ already would allow for impressive results in
cavity QED with single atoms.

By comparison, the state of the art for Fabry-Perot cavities has
already achieved the following results for the $\mathrm{TEM}_{00}$
modes~\cite{science}: a cavity finesse of $\mathcal{F}=4.8\times
10^{5}$, a mode volume of $V_{\mathrm{m}}\approx 1.69\times
10^{3}\mu \mathrm{m}^{3}$, coupling coefficient $g_{0}/(2\pi
)\approx 110\mathrm{MHz}$, saturation photon number $n_{0}\approx
2.82\times 10^{-4}$, and critical atom number $N_{0}\approx
6.13\times 10^{-3}$. If one then looks at possible limits of
Fabry-Perot technology for cavity QED as analyzed in
Ref.~\cite{hoodye}, the following may be possible; a cavity of
length $\lambda_{0} /2$ with a cavity finesse of
$\mathcal{F}=7.8\times 10^{6}$ yields coupling coefficient
$g_{0}/(2\pi )\approx 770\mathrm{MHz}$, saturation photon number
$n_{0}\approx 5.7\times 10^{-6}$, and critical atom number
$N_{0}\approx 1.9\times 10^{-4}$.

It is encouraging that the currently achievable results for small
sphere manufacture would already allow the WGMs to compete
favorably with the current state of the art in Fabry-Perot cavity
QED. However, if one were able to manufacture and couple to
spheres at the optimal size $a\approx 7.83\mu \mathrm{m}$ with a
$Q\sim 9.76\times 10^{8}$ (the theoretical maximum predicted from
the analysis of Section \ref{losses}), the following results could
be achieved: a mode volume of $V_{\vec{P}}\approx 90\mu
\mathrm{m}^{3}$, coupling coefficient $g_{a}/(2\pi )\approx
318\mathrm{MHz}$, saturation photon number $n_{0}\approx
3.36\times 10^{-5}$, and critical atom number $N_{0}\approx
9.28\times 10^{-6}$. This would represent a significant
improvement over the current Fabry-Perot technology and be
competitive with the likely limits of Fabry-Perot technology.
However, even short of achieving this stated maximum $Q$ for the
WGMs, impressive results can already be attained. With a quality
factor $Q\sim 0.8 \times 10^{7}$ at the optimal sphere radius
$a\approx 7.83\mu \mathrm{m}$, one would obtain these same results
(i.e., $g_{a}/(2\pi )\approx 318\mathrm{MHz}$ and saturation
photon number $n_{0}\approx 3.36\times 10^{-5}$), except that the
critical atom number, $N_{0}$, would increase to $N_{0}\approx
1.13\times 10^{-3}$. This is still an impressive gain over the
current capabilities of Fabry-Perot cavities for the saturation
photon number, with room for improvement in the critical atom
number.

Overall, we thus find that the technologies of microspheres and
Fabry-Perot resonators each have their advantages and
disadvantages. However, there is one notable advantage of
microspheres; they can be made cheaply and relatively simply given
sufficient training and skill. By contrast, the Fabry-Perot
cavities considered here require specialized coating runs with
expensive equipment and considerable expertise, which is to be
found at only a few locations worldwide. This alone makes
microspheres an attractive alternative to Fabry-Perot cavities for
cavity QED. Another unique advantage of the WGMs is the ability to
control the cavity decay rate, $\kappa$, by controlling the
coupling efficiency into and out of the microsphere (e.g., by
adjusting the distance between a coupling prism and the
microsphere \cite{gorodetsky3}.) Furthermore, as one moves to the
limit of small cavities, the open geometry of microspheres offers
a considerable advantage when compared to the geometry of
Fabry-Perot cavities. Such possibilities combined with our
projected values of the critical parameters, $(n_{0},N_{0})$,
shown in Fig.~\ref{fabry_micro} point to the competitiveness of
microspheres with current and future Fabry-Perot technology and
demonstrate their potential as a powerful tool for cavity QED in
the regime of strong coupling.

\begin{acknowledgments}
We thank K.~Birnbaum, S.~J.~van~Enk, C.~Hood, V.~Ilchenko,
A.~Kuzmich, R.~Legere, P.~Lodahl, T.~Lynn, H.~Mabuchi,
J.~McKeever, T.~Northup, D.~Vernooy, and J.~Ye for helpful
discussions. This work was supported by the National Science
Foundation, by the Office of Naval Research, and by the Caltech
MURI on Quantum Networks administered by the Office of Army
Research.
\end{acknowledgments}


\bibliography{micro}

\begin{thebibliography}{32}
\bibitem[{r_q()}]{r_qfootnote}
\bibinfo{note}{Because the WGMs are the modes of an open resonator, the mode
  volume, $V_{\vec{P}}$, diverges as $r_{Q}\rightarrow\infty$. However, this
  divergence is logarithmic, and $V_{\vec{P}}$ is quite insensitive to the
  choice of $r_{Q}$ for a large range of values. For example, at the optimal
  sphere size discussed in Section~\ref{strong_coupling_cesium}, where $l =
  76$, $V_{\vec{P}}$ varies by less than $1\%$ for $10<\frac{2\pi
  n}{\lambda_{0}}(r_{Q}-a)<10^{3}$. In this case, we take $(r_{Q}-a)\approx
  100\mu\mathrm{m}$}.
\end{thebibliography}

\end{document}